\documentclass[prb,  superscriptaddress, twocolumn,
aps,showpacs]{revtex4-1} 

\usepackage[dvips]{graphicx}
\usepackage[dvips]{hyperref}
\usepackage{url}
\usepackage{amsmath}

\def\beq {\begin{equation}}
\def\eeq {\end{equation}}

\def\bfk {\mathbf{k}}

\date{\today}
\begin{document}

\title{The role of non-local exchange in the electronic structure of correlated oxides}

 \author{Federico Iori}
  \affiliation{Nano-Bio Spectroscopy group and ETSF Scientific Development Centre,
  Dpto. F\'isica de Materiales, Universidad del Pa\'is Vasco,
  Centro de F\'isica de Materiales CSIC-UPV/EHU-MPC and DIPC,
  Av. Tolosa 72, E-20018 San Sebasti\'an, Spain}

\author{Matteo Gatti}
  \affiliation{Nano-Bio Spectroscopy group and ETSF Scientific Development Centre,
  Dpto. F\'isica de Materiales, Universidad del Pa\'is Vasco,
  Centro de F\'isica de Materiales CSIC-UPV/EHU-MPC and DIPC,
  Av. Tolosa 72, E-20018 San Sebasti\'an, Spain}

\author{Angel Rubio}
  \affiliation{Nano-Bio Spectroscopy group and ETSF Scientific Development Centre,
  Dpto. F\'isica de Materiales, Universidad del Pa\'is Vasco,
  Centro de F\'isica de Materiales CSIC-UPV/EHU-MPC and DIPC,
  Av. Tolosa 72, E-20018 San Sebasti\'an, Spain}
  \affiliation{Fritz-Haber-Institut der Max-Planck-Gesellschaft, 
  Theory Department, Faradayweg 4-6, D-14195 Berlin-Dahlem, Germany}

\begin{abstract}

We present a systematic study of the electronic structure of  several prototypical correlated transition-metal oxides: 
VO$_2$, V$_2$O$_3$, Ti$_2$O$_3$, LaTiO$_3$, and YTiO$_3$. 
In all these materials, in the low-temperature insulating phases   
the local and semilocal density approximations (LDA and GGA) of density-functional theory yield a metallic Kohn-Sham band structure. 
Here we  show that, without invoking strong-correlation effects, the role of non-local exchange is essential to cure the LDA/GGA 
delocalization error and provide a band-structure description of the electronic properties  in qualitative agreement with the experimental photoemission results.
To this end, we make use of hybrid functionals that mix a portion of non-local Fock exchange with the local LDA exchange-correlation potential.
Finally, we discuss the advantages and the shortcomings of using hybrid functionals for correlated transition-metal oxides.

\end{abstract} 


\maketitle

\section{Introduction}

Nowadays, the standard model of electronic structure calculations is
based on density-functional theory (DFT) 
in the Kohn-Sham (KS) formalism \cite{lda}.
The DFT-KS scheme, also in its simplest approximations like the local-density approximation (LDA) \cite{lda} 
or the generalized-gradient approximation (GGA) \cite{pbe},  is generally highly successful in a very large variety of applications.
Thus, when a LDA (or GGA) KS band structure turns out to be metallic in an insulating compound (e.g. in a transition-metal oxide), 
the result is often interpreted as a direct indication for strong electron correlation effects in the material, 
and a failure of the band-structure picture. 
A possible strategy that has been followed in the literature to overcome these difficulties is to resort to model approaches, 
like the multiband Hubbard model. In the LDA+U approach \cite{ldau},
LDA band structures are supplemented by an on-site Coulomb interaction (the Hubbard U) acting 
only on the ``correlated'' subset of the electronic degrees of freedom. In a higher level of theory, 
dynamical mean-field theory (DMFT) \cite{ldadmft}, the Hubbard model is 
further mapped onto an Anderson impurity model, which can be then solved with different techniques, allowing for the description of dynamical effects beyond LDA+U. 
In those cases LDA is claimed to be inadequate to capture the strong interactions taking place between correlated  electrons in partially filled $d$ (or $f$) shells, which  give rise to narrow bands in the solid.

However, we remark that DFT is a ground-state theory and KS band structures are not meant to describe the electronic excitations measured in photoemission, 
which also define  the fundamental band gap of an insulator. 
Moreover, LDA lacks the derivative discontinuity of the local KS exchange-correlation (xc) potential \cite{sham83} $V_{xc}(r)$ and suffers 
from a severe delocalization error \cite{sanchez08}, which is particularly relevant for localised $d$ and $f$ electrons.
The underestimation of the fundamental band gaps in $sp$ semiconductors is well known\cite{onida02,schilfgaarde06}, and understood in terms of self-energy corrections at the GW level of approximation \cite{hedin65,hybersten86}.
This underestimation sometimes may lead to metallic band structures also in ``weakly correlated''  small-gap semiconductors, like in germanium.  

Here we consider several prototypical correlated transition-metal oxides that have been studied by other methods in the past years: VO$_2$, V$_2$O$_3$, Ti$_2$O$_3$, LaTiO$_3$, and YTiO$_3$. 
In all the low-temperature insulating phases of these materials, 
KS-LDA yields a metallic band structure. We thus address the following question:
Is this just a result of the inadequacy of LDA to deal with strong correlations or, rather, is this finding related to the systematic KS-LDA underestimation of  band gaps that occurs also in ``weakly correlated'' semiconductors and get enhanced in the oxides?
To answer this question we make use of a generalized Kohn-Sham (gKS) scheme \cite{gks}, where the local KS xc potential $V_{xc}(r)$ is replaced by
a spatially non-local  $V_{xc}(r,r')$. In these hybrid functionals, the non-local Fock exchange potential is mixed
with the local (LDA or GGA) KS xc potential \cite{bylander90,pbe0,hse}. In these approaches,  
the difference with LDA results stems solely from the non-local exchange term. This is not supposed to improve the description of electronic correlation. 
Therefore, if strong correlations are responsible for invalidating the KS-LDA description of these insulators, then one should expect to find the same problems in the gKS scheme. We will instead show how, without invoking strong-correlation effects, the role of non-local exchange is essential to cure most of the LDA delocalization error and provide a band-structure description of the electronic properties of several transition-metal oxides in qualitative agreement with the experimental results.

The paper is organized as follows. In Sec. \ref{sec2} we briefly introduce the generalized Kohn-Sham (gKS) scheme and the hybrid-functional parametrization that we use, and compare them with standard methods that treat electronic correlations. In Sec. \ref{sec3}  we present and discuss the results that we have obtained for the various transition-metal oxides, also across their metal-insulator phase transitions (MIT) \cite{imada98}. Finally, in Sec. \ref{sec4} we draw our conclusions on the basis of these results and discuss the advantages and shortcomings of the use of hybrid functionals in the description of the electronic properties of correlated oxides.

\section{Method}
\label{sec2}

In the gKS scheme the xc potential is generalized to be non-local in space\cite{gks}, contrary to ordinary KS where it is local. 
In this framework, a common choice is to write the gKS non-local potential $V_{xc}(r,r')$ as a sum of a non-local exchange term and a local KS correlation potential\cite{bylander90}: $V_{xc}(r,r')=V_x(r,r')+V_c(r)$. For $V_c(r)$ a LDA or GGA expression is adopted, while the non-local exchange term $V_x(r,r')$ is obtained mixing the KS local exchange $V_x(r)$ with a fraction $\alpha$ of the non-local Fock operator $V_x^{F}(r,r') = -\gamma(r,r')v(r,r')$, built using gKS orbitals in the one-particle density matrix $\gamma$ [$v(r,r')= 1/{\mid}r-r'\mid$ is the Coulomb interaction]:
\beq
V_x(r,r') = \alpha V_x^{F}(r,r') + (1-\alpha)V_x(r).
\label{vxnl}
\eeq
This construction is often justified from considerations based on the adiabatic-connection formula, which fix the mixing parameter to be $\alpha=0.25$.\cite{pbe0} 
The Coulomb interaction can be then split into a sum of a long-range and a short-range term $v(r,r') = v^{LR}(r,r') + v^{SR}(r,r')$, respectively as:
\beq
v(r,r') =  \frac{\text{erf}(\mu{\mid}r-r'\mid)}{{\mid}r-r'\mid} + \frac{1- \text{erf}(\mu{\mid}r-r'\mid)}{{\mid}r-r'\mid}.
\eeq
Here both the choice of the  $\text{erf}$ error function and the parameter $\mu$ are arbitrary.
Replacing the bare interaction $v$ by either term, the same separation is obtained in the local and non-local exchange potentials appearing in  \eqref{vxnl}: $V_x^{F}(r,r') = V_x^{F,LR}(r,r')+V_x^{F,SR}(r,r')$ and $V_x(r) = V_x^{LR}(r)+V_x^{SR}(r)$.
Assuming that the effects of $V_x^{F,LR}(r,r')$ and $V_x^{LR}(r)$ compensate each other, the following approximation for \eqref{vxnl} is introduced:
\beq
V_x(r,r') = \alpha V_x^{F,SR}(r,r') + (1-\alpha)V_x^{SR}(r)+ V_x^{LR}(r).
\label{vxnlapprox}
\eeq
Regrouping the various terms contributing to the xc gKS potential, one finally finds:
\beq
V_{xc}(r,r') = \alpha [V_x^{F,SR}(r,r') - V_x^{SR}(r)] + V_{xc}(r),
\label{vxcnl}
\eeq
where the correction to the KS local potential $V_{xc}(r)$ (with the original Coulomb interaction) stems entirely from the first term in the r.h.s.
The final approximation depends on the two parameters $\alpha$ and $\mu$.
With $\mu=0$, $\alpha=0.25$, and using the PBE GGA xc potential for the local part  $V_{xc}(r)$, one finds the PBE0 approximation.\cite{pbe0}
In the HSE06 hybrid functional\cite{hse}, instead, the value of $\mu=0.2 \text{ \AA}^{-1}$ is obtained by numerically fitting the results against a benchmark set of data. 

More in general, both the $\alpha$ and $\mu$ parameters play the role of effective screening of the Coulomb interaction \cite{marques11}. The non-local potential \eqref{vxnl} with $\mu=0$ can be alternatively understood as a static approximation to the many-body GW self-energy\cite{hedin65}. By identifying $1/\alpha$ as an effective static dielectric constant $\epsilon$, $\alpha V_x^{F}$  can be seen  as a screened exchange potential, while the local part of \eqref{vxnl} acts as an approximation to the Coulomb hole term.\cite{hedin65,gygi89}
In fact, varying $\alpha$ between 0 and 1 in Eq. \eqref{vxnl}, one in practice interpolates between the KS underestimation and the Hartree-Fock (HF) overestimation of band gaps, with the possibility to get close to the experimental results.
Moreover, the use of a finite value for $\mu$, together with neglecting the corresponding long-range terms in \eqref{vxnlapprox}, efficiently acts as a further screening of the Coulomb interaction.\cite{hse,krukau} By increasing the value of $\mu$, the two-point distance ${\mid}r-r'\mid$ beyond which the Coulomb interaction is cut off becomes shorter. In fact, screening the long-range Coulomb interaction is crucial for many properties in bulk materials \cite{bruneval_jcp}, especially for small-gap and metallic systems. While for $\mu=\infty$ Eq. \eqref{vxnlapprox} reduces to the KS local potential $V_{xc}(r)$, a finite $\mu$ tunes the correction to KS stemming from the difference between non-local Fock and local exchange terms in  \eqref{vxnlapprox}. For fixed $\alpha$, increasing $\mu$ gives less weight to this correction.

While LDA suffers from a delocalization error, HF instead is affected by an excess of localization\cite{sanchez08}. 
Hence, the inclusion in the functional \eqref{vxcnl} of a partial contribution of non-local Fock exchange leads to a localization correction with respect to LDA. 
This affects in particular $d$ and $f$ states, which are more localised and generally suffer from a self-interaction error more than $s$ and $p$ states.\cite{sanchez06} 
Even though it is derived from a very different point of view, this effect of localization of $d$ and $f$ orbitals is shared also by the LDA+U approach. As we will show in the following, curing the delocalization error of LDA is the key to get results in better qualitative agreement with experiment.
On the other hand, both LDA+U (as well as LDA+DMFT) and the hybrid functional  \eqref{vxcnl} depend on parameters (like the Hubbard U and J  in the former 
and $\alpha$ and $\mu$ in the latter), which limits their predictivity power. However, in the hybrid functional \eqref{vxcnl} there is no need of an additional term for correcting the (unknown) spurious double counting, while it is case for LDA+U or LDA+DMFT. 
Moreover,  in the hybrid functional all the electrons are treated on equal footing. 

Modeling efficiently the static screening of the Coulomb interaction without introducing adjustable parameters would help to improve greatly the gKS scheme. At the same time, this should keep its computational cost cheaper than the more sophisticated many-body GW approximation, where the dynamical screening is explicitly calculated (in the random-phase approximation).
In any case, being a static approximation to the many-body self-energy, the $V_{xc}(r,r')$ functional in \eqref{vxcnl} cannot account for dynamical correlation effects, which for instance give rise to satellites in the photoemission spectra \cite{guzzo11}, and are instead accounted for, though in different manners, by both the GW approximation and the LDA+DMFT approach.

In the following we will use the hybrid functional form  \eqref{vxcnl}, as implemented in the \textsc{vasp} computer code\cite{vasp,hsevasp}.
For the local $V_{xc}(r)$ part we adopt a LDA xc potential. In the case of VO$_2$ we will discuss the effect of different choices of the two parameters $\alpha$ and $\mu$, while for the rest of the paper we will fix $\alpha=0.25$ and $\mu=0.2 \text{ \AA}^{-1}$ as in the HSE06 functional.

\section{Results and discussion}
\label{sec3}
      
\subsection{VO$_2$}
\label{sec_vo2}

VO$_2$ undergoes a twofold phase transition at 340 K\cite{morin59}. The MIT is accompanied by a lowering of the symmetry of the crystal structure, from rutile to monoclinic, with a doubling of the unit cell and a dimerization of V atoms along the rutile $c$ axis. 
It has been long debated\cite{zylbersztejn75,wentzcovitch94} which of the two aspects, the electronic or the structural change,  is the key to drive the phase transition. 

In VO$_2$ both LDA\cite{eyert02} and standard single-site DMFT\cite{liebsch05,laad06} are unable to get the insulating band gap, 
while LDA+U \cite{liebsch05,korotin02} has problems with the metallic phase and gives an ordered magnetic phase for the insulator, 
contrary to the experiment. The deficiencies of single-site DMFT have been corrected by its extension to  cluster DMFT\cite{biermann05,lazarovitis10}, where the local impurity is taken to be a V dimer instead of a V atom, as in single-site DMFT.
On the other hand, parameter-free GW calculations\cite{gatti07} have shown that in the insulating phase KS-LDA wavefunctions are not a sufficiently good approximation to quasiparticle (QP) wavefunctions. The LDA error is due to an excessive delocalization of the KS wavefunctions at the Fermi level. They turn out to be too isotropic, underestimating the effect of the V dimerization along the $c$ axis\cite{wentzcovitch94}, 
and the corresponding bonding-antibonding splitting of the V $a_{1g}$ states. 
Once this LDA error is corrected by using better QP wavefunctions, as obtained in a restricted self-consistent GW scheme, the results correctly reproduce the electronic properties of both phases  \cite{gatti07,continenza99,sakuma08} and also show that the satellite in photoemission spectrum of the metallic phase is related to a neutral (plasmon) excitation visible in the loss function  \cite{gatti11}.
\begin{figure}
\begin{center}
\includegraphics[width=\columnwidth]{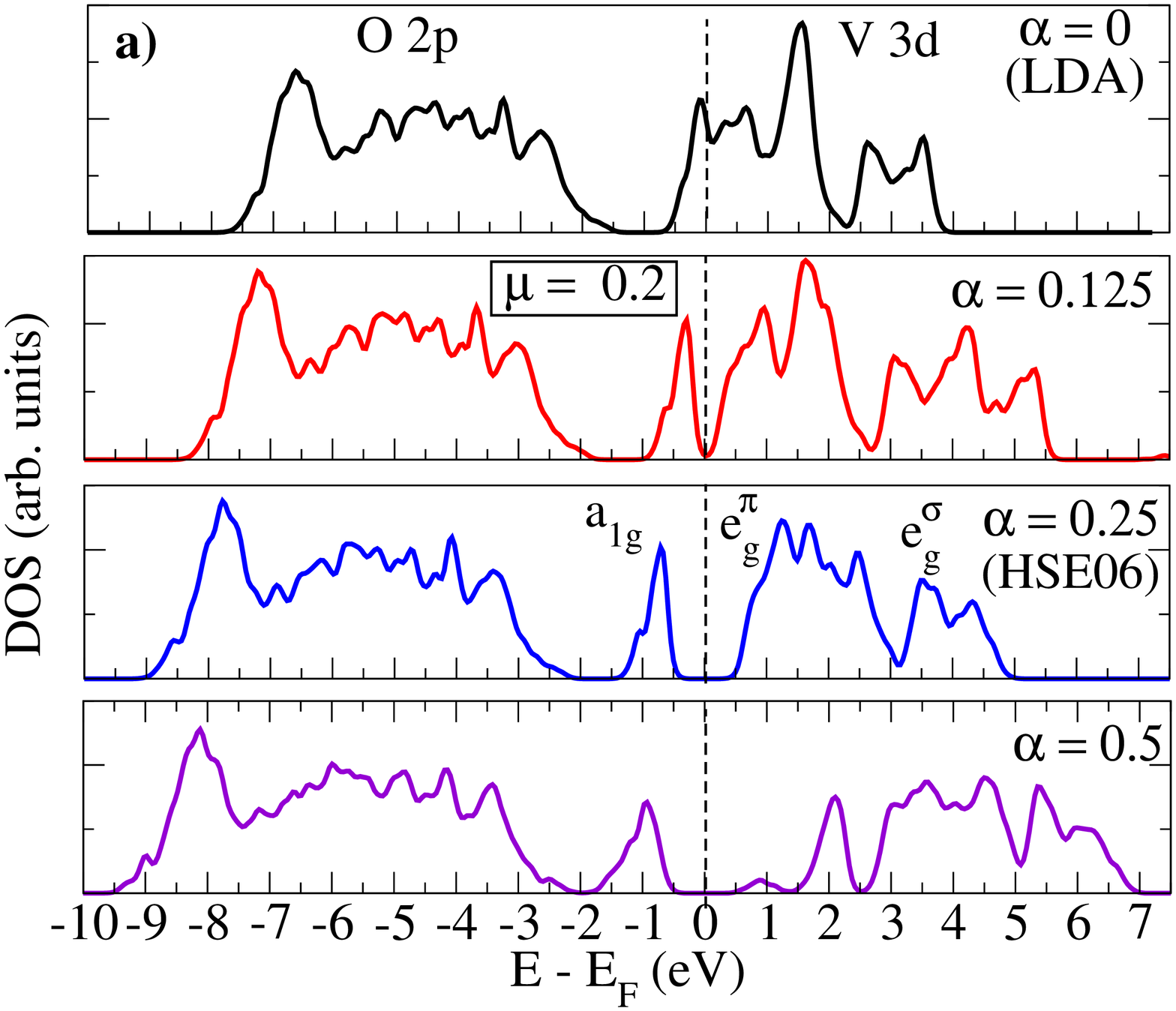}\\[0.2cm]
\includegraphics[width=\columnwidth]{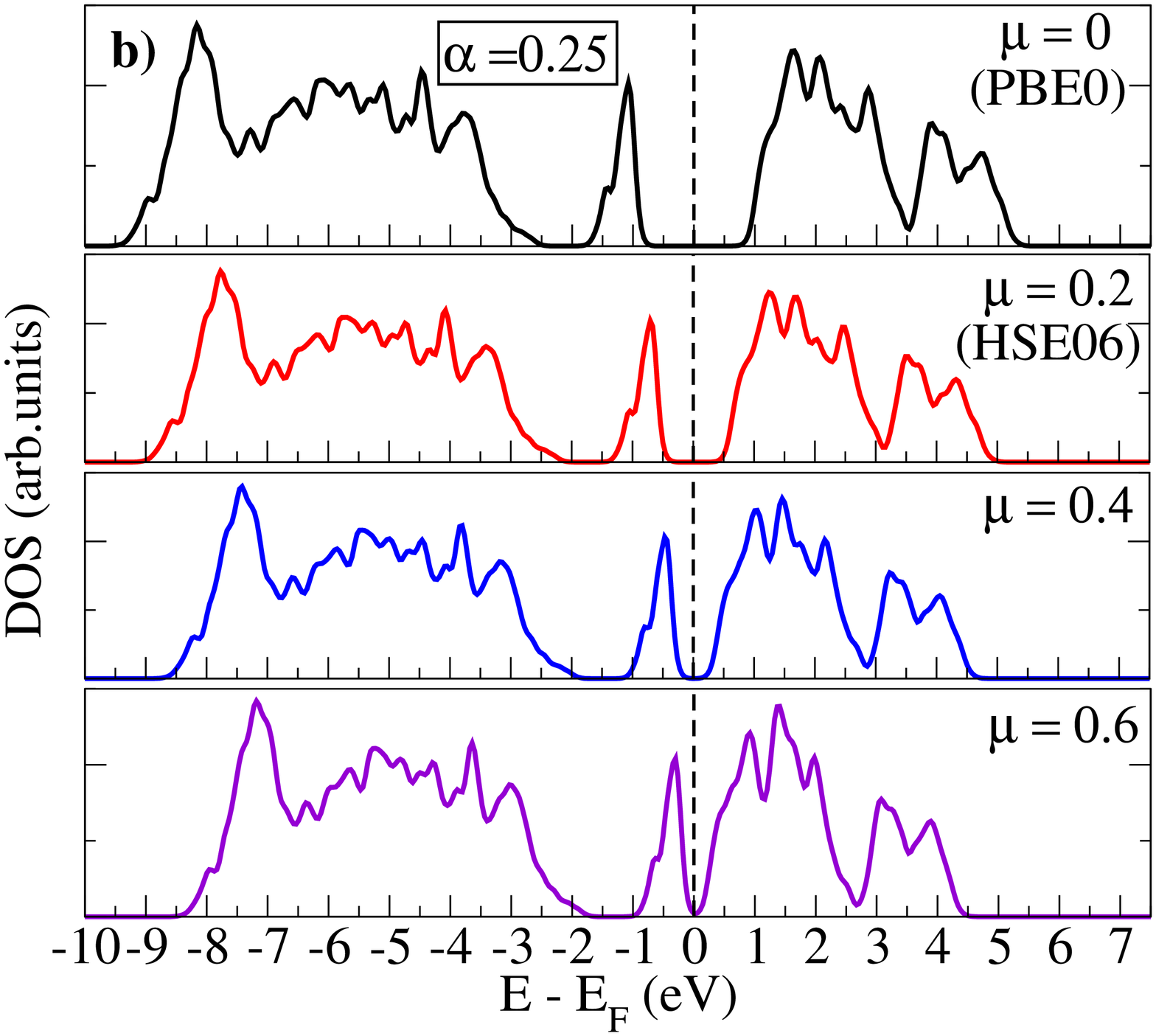}
\caption{(Color online) Densities of states of insulating VO$_2$ obtained with the hybrid functional of Eq. \protect\eqref{vxcnl} according to different choices of the mixing parameter $\alpha$ and the screening parameter $\mu$. In (a) the topmost panel corresponds to LDA. In the other panels $\mu=0.2$ \AA$^{-1}$  and $\alpha$ is raised up to 0.5. In (b) $\alpha$ is fixed to 0.25 and the results are obtained for different values of $\mu$. In the topmost panel the PBE0 result is retrieved. Here, and in all the following figures, the Fermi energy for insulators is set in the midpoint of the band gap.}
\label{figvo2_1}
\end{center}
\end{figure}

\begin{center}
\begin{table}
\caption{Values of the fundamental band gap, the O $p$ bandwidth, the O $p$ - V $a_{1g}$ separation, and the V $a_{1g}$ bandwidth in insulating VO$_2$ depending on the different choices for the $\alpha$ and $\mu$ parameters in the hybrid functional of Eq. \protect\eqref{vxcnl}.}
\begin{tabular}{c c c c c c}
\hline
\hline
$\alpha$ & $\mu$ & Band gap &  O $p$ &  O$p$-V$a_{1g}$ & V$a_{1g}$ \\
         & (\AA$^{-1}$) & (eV) & (eV) & (eV) & (eV)\\
\hline
 0 & -     &     0.00    & 6.23   & 0.96  & -\\
 0.125 & 0.2     &  0.00  & 6.54  & 1.02  & 0.60\\
 0.250 & 0.2     &  1.13  & 6.85  & 0.95 & 0.63\\
 0.500 & 0.2     &  1.17  & 7.58  & 0.09 & 1.16\\
\hline
 0.250 & 0.0     &  1.88  & 6.86  & 0.94 & 0.64\\
 0.250 & 0.4     &  0.65  & 6.74  & 0.96 & 0.60\\
 0.250 & 0.6     &  0.00  & 6.62  & 0.97 & 0.59\\
\hline
\hline
\end{tabular}
\label{tabvo2}
\end{table}
\end{center}

A very recent calculation\cite{eyert11} within the HSE06 hybrid functional 
obtained a gap in density of states (DOS) of the insulator, also concluding in favor of the structural distortion as the key to explain the insulating gap.
Here, more in detail, we start our investigation by analysing the performance of the hybrid functional \eqref{vxcnl} according to different choices of the values of the mixing parameter $\alpha$ and the screening parameter $\mu$. 
We use the experimental crystal structures for the two phases\cite{mcwhan74,longo70}, a 6$\times$6$\times$6 $\bfk$-point grid for the insulator and a 12$\times$12$\times$12 one for the metal.
As we can see in Fig. \ref{figvo2_1}, the hybrid functional \eqref{vxcnl} correctly yields a gap in the insulating phase of VO$_2$ for many choices of the two parameters, correcting the LDA error [which is retrieved for $\alpha=0$, see topmost panel in Fig. \ref{figvo2_1}(a)].
On the other hand, the DOS is highly sensitive to the values of $\alpha$ and $\mu$, both for the size of the band gap and for the separation between the highest occupied V $a_{1g}$ state and the rest of the O $p$ states at higher binding energies (see Tab. \ref{tabvo2}). Either reducing $\alpha$ [Fig. \ref{figvo2_1}(a)] or increasing $\mu$ [Fig. \ref{figvo2_1}(b)], the gap decreases until it disappears for $\alpha < 0.125$ or $\mu > 0.6$ \AA$^{-1}$.
Keeping $\alpha$ fixed, the reduction of $\mu$ from 0.6 to 0 {\AA}$^{-1}$  has mainly the effect of a rigid expansion of the DOS [see Fig. \ref{figvo2_1}(b)]. 
On the contrary, fixing $\mu$ and changing $\alpha$ leads also to larger modifications in the shape of all the structures appearing in the DOS, both for occupied and the unoccupied states [see Fig. \ref{figvo2_1}(a)].
Thus,  the addition of non-local exchange to the KS-LDA functional leads to an improved qualitative agreement with  experiment, i.e. a sizable band gap is obtained, without the need to invoke strong correlation effects.
However, the comparison with photoemission spectra depends quantitatively on the choice that one makes for the values of the two parameters $\alpha$ and $\mu$. In fact, they both physically act as screening of the Coulomb interaction. 
However, their fine tuning generally depends on an adequate microscopic description of the screening in the actual material.

\begin{figure}
\begin{center}
\includegraphics[width=\columnwidth]{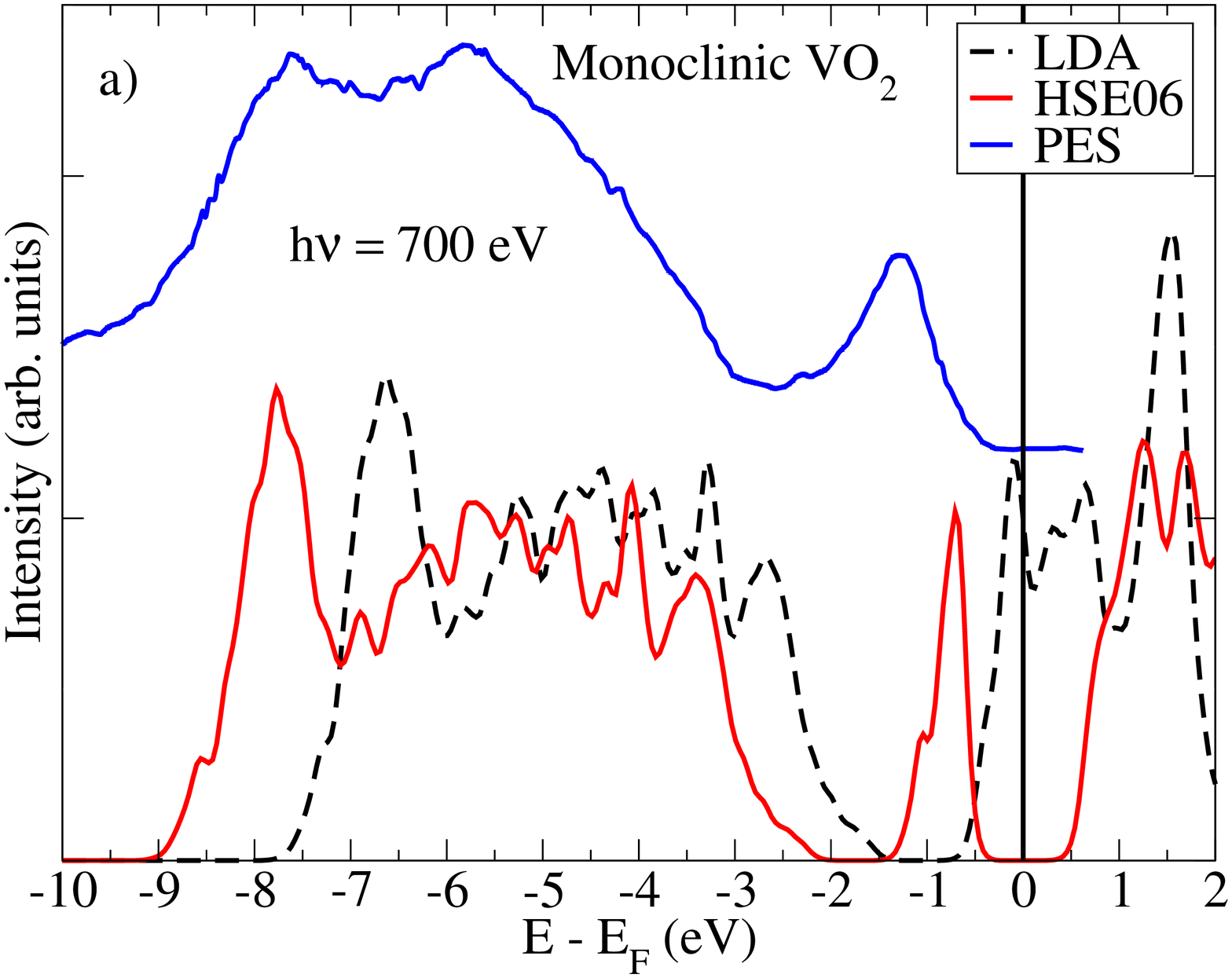}\\[0.2cm]
\includegraphics[width=\columnwidth]{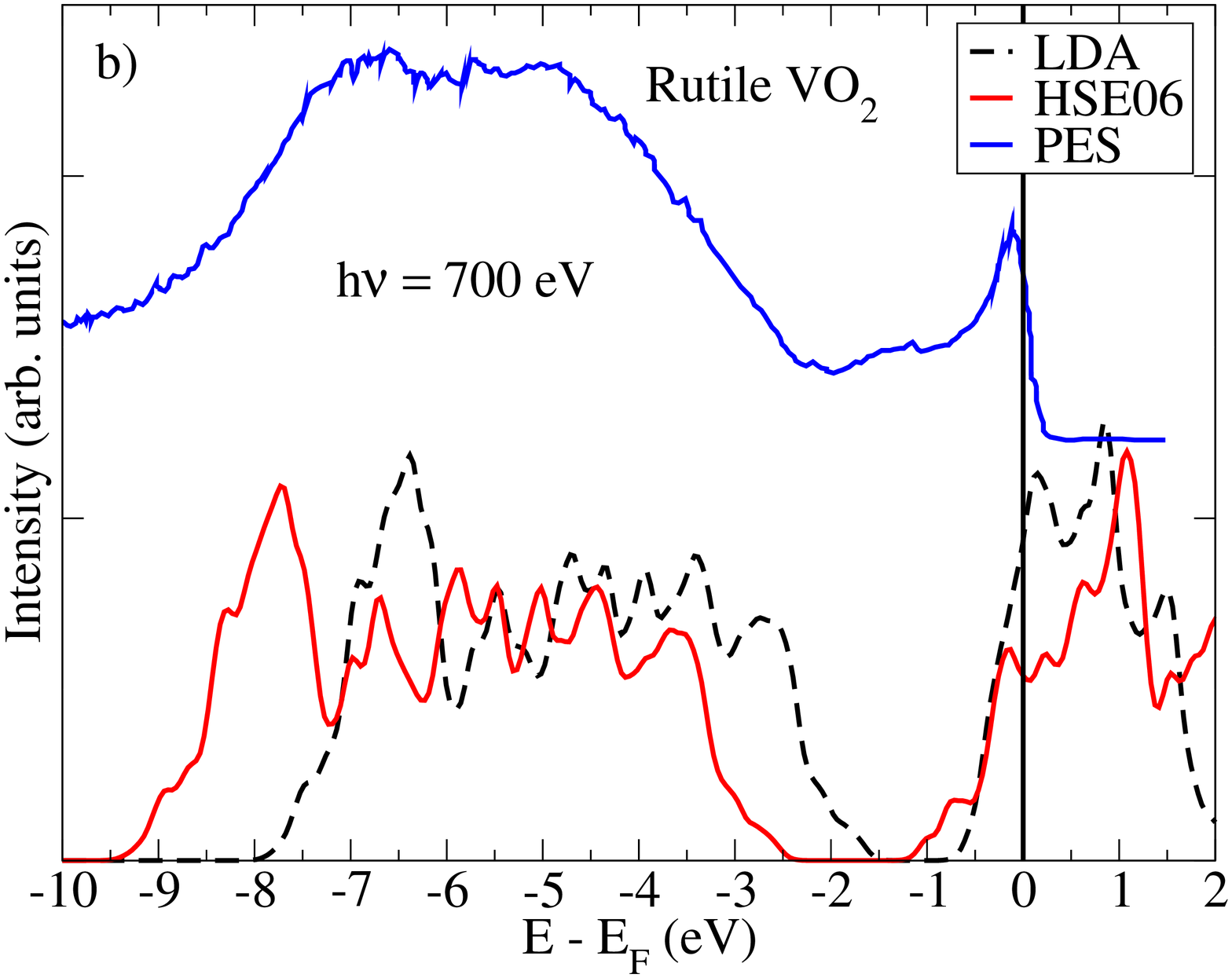}
\caption{(Color online) Comparison of the calculated density of states of insulating monoclinic (upper panel) and metallic rutile (bottom panel) phases of VO$_2$ with the experimental photoemission spectra from Ref. \protect\cite{koethe06}. The MIT is correctly reproduced by the HSE06 hybrid functional, while the LDA DOS is always metallic.}
\label{figvo2_2}
\end{center}
\end{figure}

In Fig. \ref{figvo2_2} we compare the calculates DOS for both phases of VO$_2$ with the experimental photoemission spectra\cite{koethe06}. 
When comparing with experiments, it is essential to have bulk-sensitive photoemission data, which can be obtained by using high-energy photons as in a X ray photoemission (XPS) or, even better, a hard X ray photoemission (HAXPES) setup\cite{haxpes}. In fact, at low photon energies, photoemission spectroscopy is mainly surface sensitive. However, in these materials the electronic properties of the surfaces are generally different from the bulk. 
Here, and in the rest of paper, we use $\mu = 0.2$ {\AA}$^{-1}$ and $\alpha = 0.25$, in agreement with the HSE06 parametrization.
With this choice we find a gap of 1.13 eV, in excellent agreement with the value of 1.1 eV from Ref. \cite{eyert11}, where a GGA instead of LDA KS local functional has been used though. This shows that using either LDA or GGA KS functional does not have an influence on the DOS.
Both results overestimate the experimental band gap, which is 0.6 eV in the insulator \cite{koethe06}.
Nevertheless, the HSE06  hybrid functional is  able to describe correctly the MIT, contrary to LDA, LDA+U, and LDA+DMFT.

\subsection{V$_2$O$_3$}

Like VO$_2$, V$_2$O$_3$ is a time-honored correlated material, whose great interest is due to its temperature-induced MIT \cite{imada98,mcwhan73}.
At T $>$ 154 K it is a paramagnetic metal, while at low temperature it becomes an antiferromagnetic insulator and undergoes a crystal distortion from a corundum to a monoclinic structure. The phase diagram is made more complex by doping with Cr, which induces a different isostructural MIT to a paramagnetic phase.
The antiferromagnetic phase has been studied in LDA+U\cite{ezhov99} (also followed by a perturbative GW calculation \cite{kobayashi08}) and LDA+DMFT\cite{panaccione06}, while the spectral properties of the metallic phase have been extensively analysed in LDA+DMFT \cite{panaccione06,mo03,keller04,poteryaev07} and in GW \cite{papalazarou09}.

\begin{figure}
\begin{center}
\includegraphics[width=0.9\columnwidth]{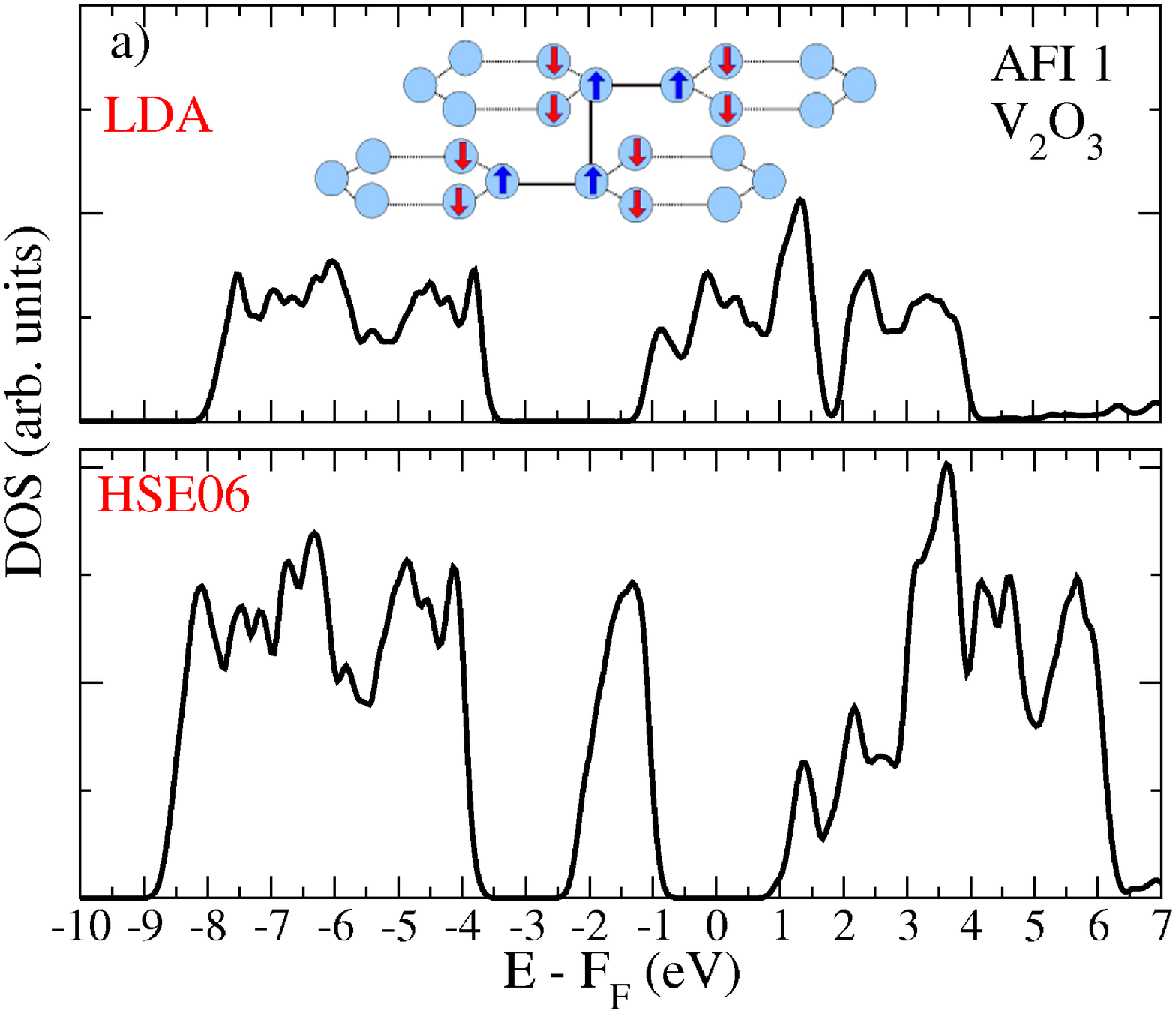}\\[0.2cm]
\includegraphics[width=0.9\columnwidth]{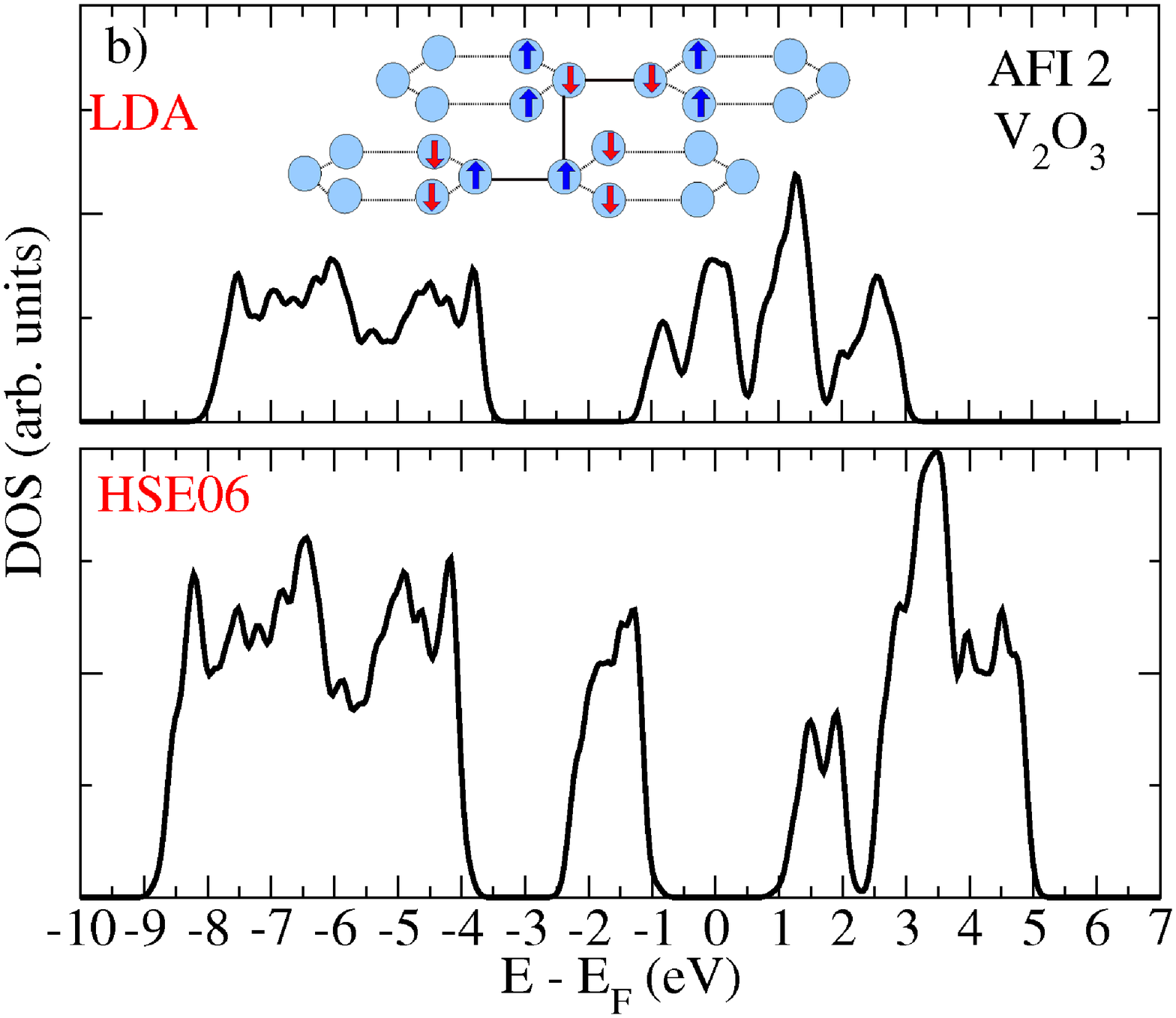}\\[0.2cm]
\includegraphics[width=0.9\columnwidth]{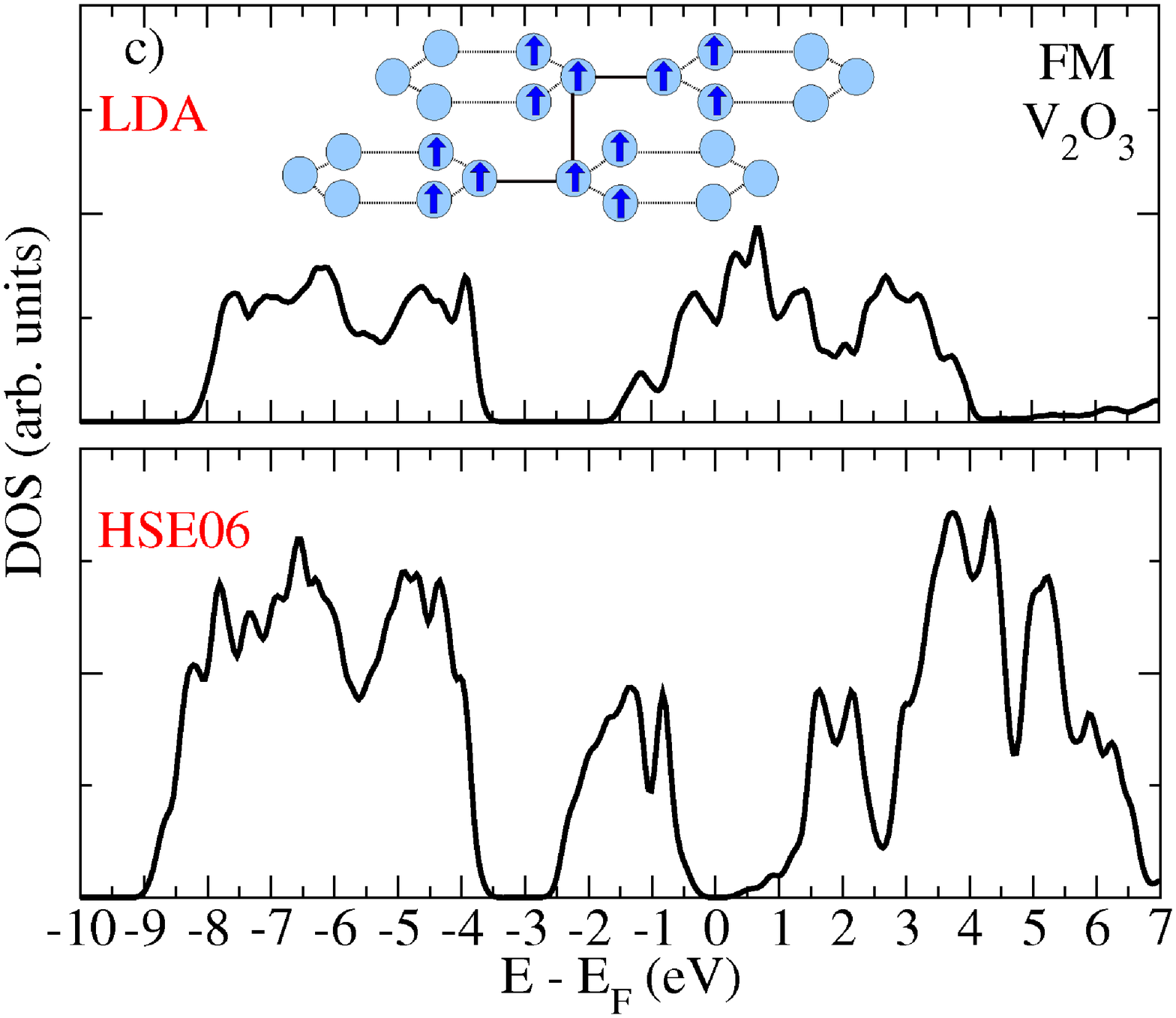}
\caption{(Color online) Density of states of monoclinic insulating V$_2$O$_3$, calculated both in LDA and HSE06, according to the different considered magnetic structures. The magnetic order is visualized in the insets to each panel, where the light blue circles schematically represent V atoms in the (distorted) hexagonal planes that characterise the monoclinic crystal structure and the arrows display the direction of the local magnetic moments.}
\label{figv2o3_1}
\end{center}
\end{figure}

Here we take the experimental lattice  parameters of the pure compounds \cite{dernier70,dernier70b}. 
We consider different magnetic configurations for the insulating phase, in order to analyse their influence on the electronic structure and compare hybrid functional results  with those found in LDA+U \cite{ezhov99}.
In the calculation we used a  6$\times$6$\times$6  grid of $\bfk$ points for the insulator, which becomes 10$\times$10$\times$10 for the metal.
In the experimental magnetic structure (AFI1) \cite{moon70}, each V atom  has one spin-parallel  neighbor and two spin-antiparallel neighbors in the (distorted) hexagonal planes of the monoclinic crystal, while the coupling between neighbors in different planes is ferromagnetic [see inset to Fig. \ref{figv2o3_1}(a)]. We also considered a ferromagnetic (FM) order and an alternative antiferromagnetic (AFI2) order, in which, with respect to experiment, the interplane magnetic coupling is inverted, while is unchanged inside the planes  [see inset to Fig. \ref{figv2o3_1}(b)].
LDA always yields a metal, regardless of the magnetic structures. Instead, HSE06 gives an insulator for all the magnetic configurations considered.  The band gap is 1.80 eV in AFI1, much larger than the 0.66 eV experimental optical gap \cite{thomas94}. 
The two antiferromagnetic DOS turn out to be very similar, while for the FM the DOS [see  Fig. \ref{figv2o3_1}(c)] is quite different and the band gap reduces to 0.7 eV. In the LDA+U calculation \cite{ezhov99}, the FM DOS is half-metallic, and the experimental antiferromagnetic structure has a 0.7 eV gap. HSE06 consistently overestimates the band gap in all the magnetic structures. This result is connected to an overestimation of the local magnetic moment, which in HSE06 is 1.8 $\mu_B$ for the AFI structures and 1.9 $\mu_B$ for the FM, whereas experimentally it is  1.2 $\mu_B$ per V atom. 
Similarly to LDA+U, the ground-state total energy difference between different magnetic structures is rather small, with the experimental AFI1 structure being the one with lowest total energy.

\begin{figure}
\begin{center}
\includegraphics[width=0.9\columnwidth]{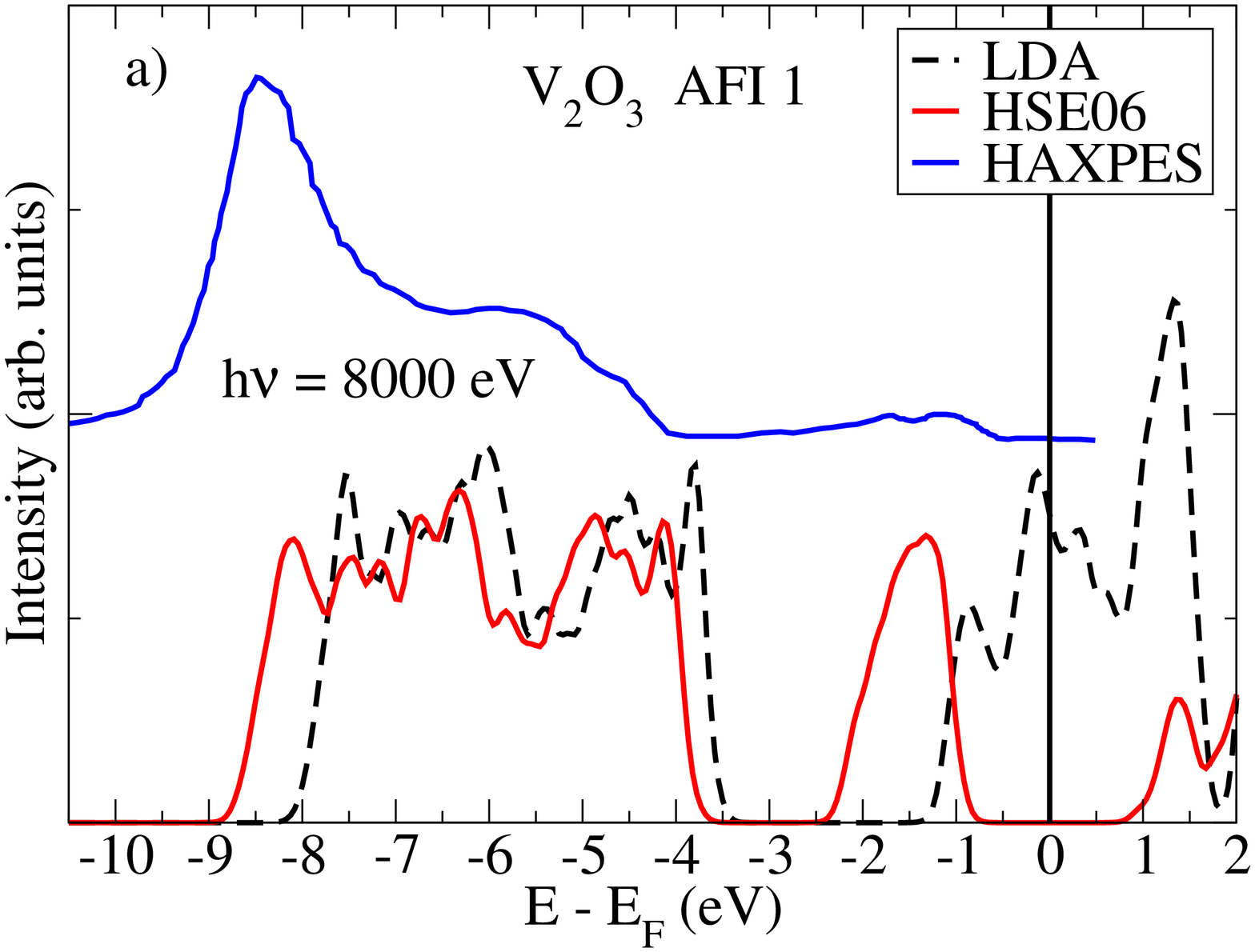}\\[1cm]
\includegraphics[width=0.9\columnwidth]{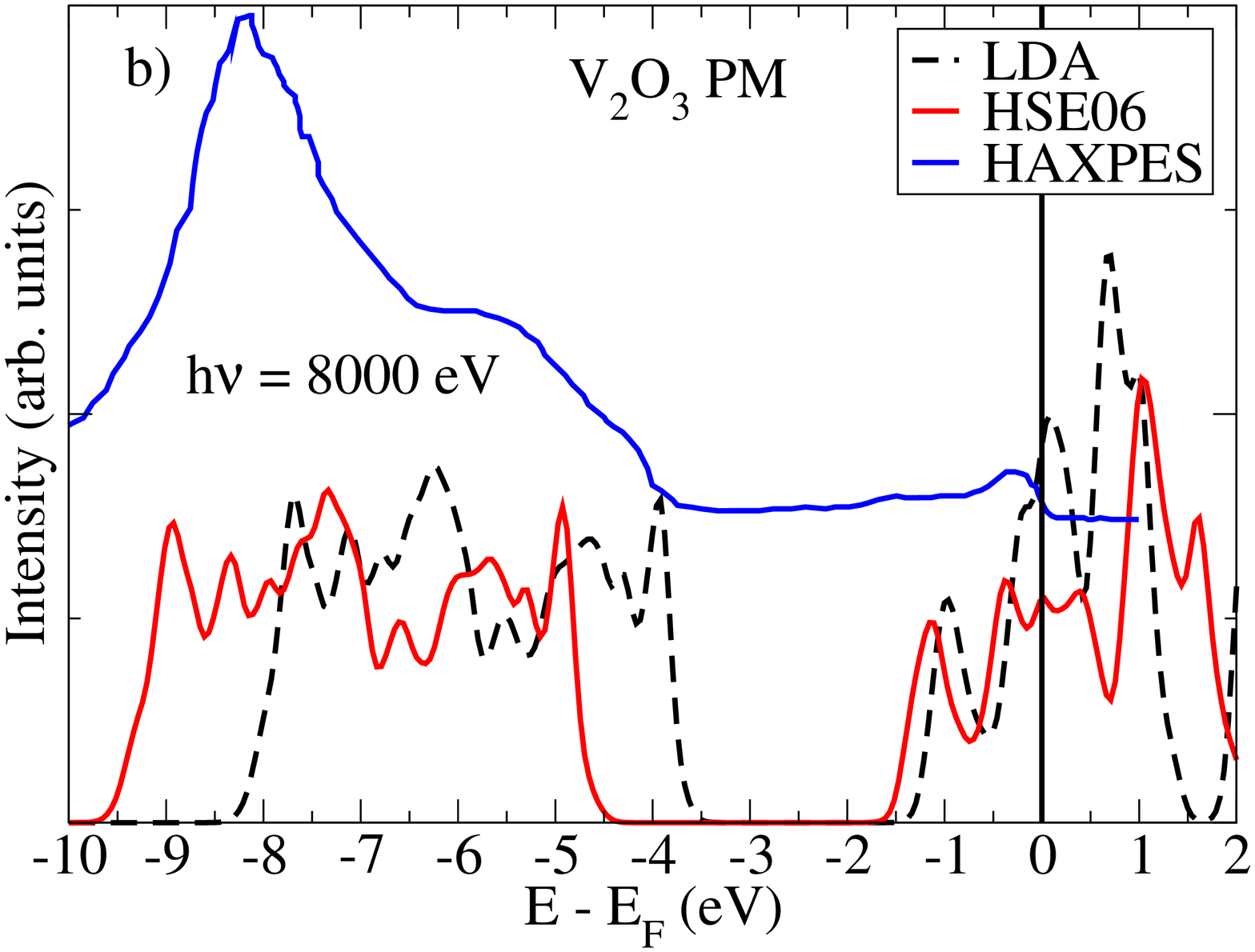}
\caption{(Color online) Comparison between the DOS calculated in LDA and HSE06 and the experimental photoemission spectra from Ref. \protect\cite{fujiwara11}. (a) Antiferromagnetic insulating phase and (b) paramagnetic metallic phases of V$_2$O$_3$.}
\label{figv2o3_2}
\end{center}
\end{figure}

In Fig. \ref{figv2o3_2} we compare the calculated DOS with the experimental HAXPES results from Ref. \cite{fujiwara11}.
As in VO$_2$ (see Sec. \ref{sec_vo2}), HSE06 is able to reproduce the MIT, contrary to LDA. 
At the same time, it also corrects the LDA underestimation of the binding energy of the O $p$ states in both phases. This is a clear illustration of the advantage of treating all the electrons on equal footing (in LDA+U or LDA+DMFT, instead, the position of O $p$ states is essentially the same as in LDA).
However, HSE06 overestimates the band width of the top valence $a_{1g}$ states, and the band gap in the insulator.

\subsection{Ti$_2$O$_3$}

Ti$_2$O$_3$ has a corundum crystal structure as metallic V$_2$O$_3$ and a formal $d^1$ configuration as VO$_2$.
Below 400 K it is insulating and above 600 K it is metallic, undergoing a very broad temperature-induced MIT\cite{morin59,honig68}.
Moreover, neither phase is magnetically ordered\cite{moon69}, and, contrary to both VO$_2$ and V$_2$O$_3$, the MIT in Ti$_2$O$_3$ is isostructural\cite{rice77}.
In fact, by raising the temperature only an increase of the $c/a$ ratio is observed, accompanied by an increase of the Ti dimer distance along the $c$ axis. 
Here we consider two crystal structures with lattice parameters measured at 296 K and 868 K \cite{rice77}, where Ti$_2$O$_3$ is insulating and metallic, respectively. 
In our calculations we used a 6$\times$6$\times$6 $\bfk$-point grid for the insulating phase and a 8$\times$8$\times$8 one for the metal.

The presence of Ti dimers along the $c$ axis leads to a large bonding-antibonding splitting of the $a_{1g}$ states, as in the insulating VO$_2$ case. Here, without additional structural changes, we can directly relate the shortening of Ti dimer distance with the increase of the $a_{1g}$ bonding-antibonding splitting, and hence the band-gap opening between the bonding $a_{1g}$ and the $e^\pi_g$ states in $t_{2g}$ subband of Ti 3$d$ orbitals. 
Also in the present case, this effect is masked in LDA by its underestimation of the anisotropy introduced by the Ti dimers.
For this reason, the LDA DOS remains metallic also at low temperatures\cite{mattheis96, eyert05}.
The relevance of the Ti dimers is underlined also by the fact that, as in VO$_2$, single-site DMFT is not able to obtain the insulating phase, 
which is instead reproduced by a cluster DMFT calculation\cite{poteryaev03} for which the local impurity is given by the pair of Ti atoms.

\begin{figure}
\begin{center}
\includegraphics[width=0.9\columnwidth]{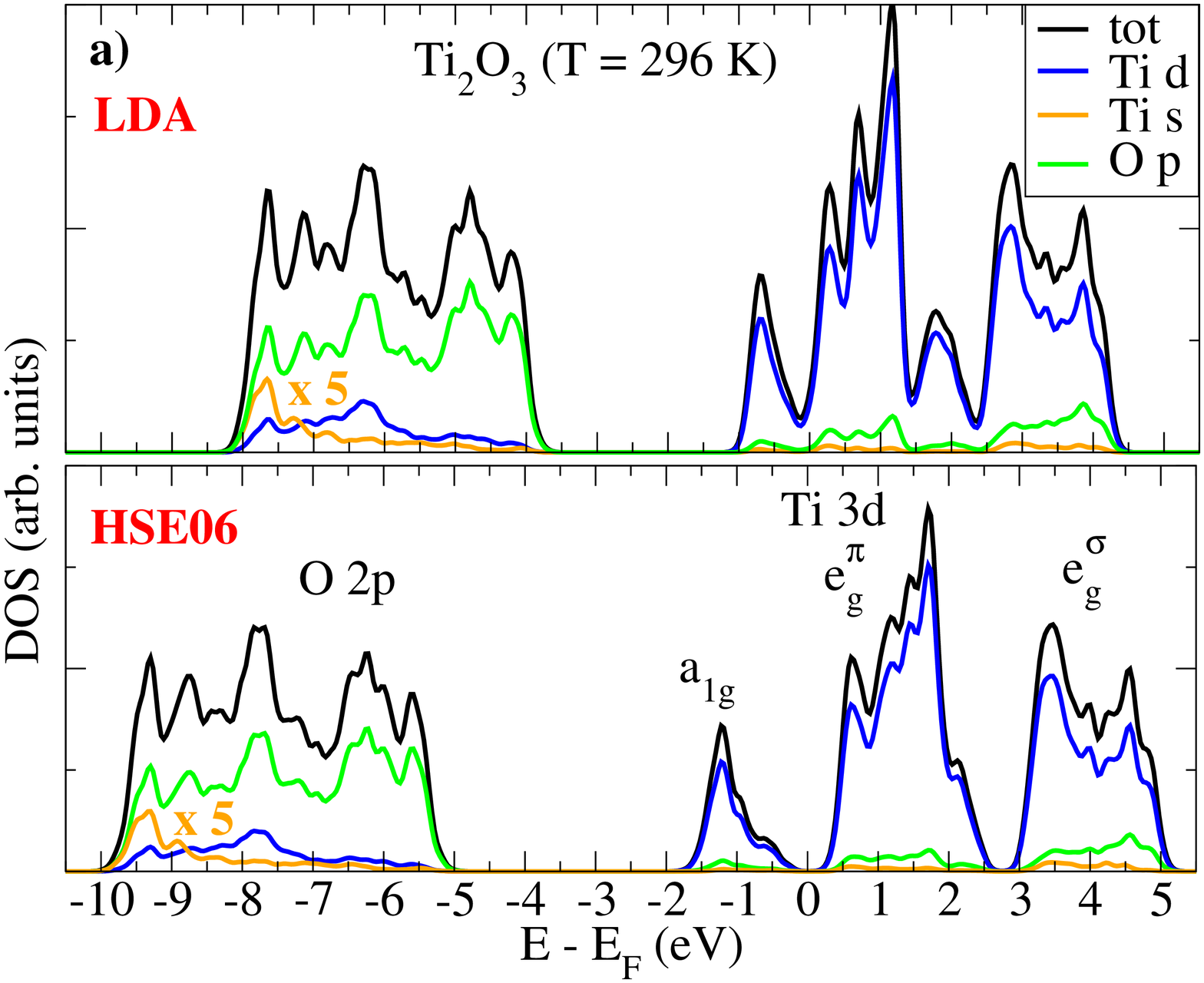}\\[1cm]
\includegraphics[width=0.9\columnwidth]{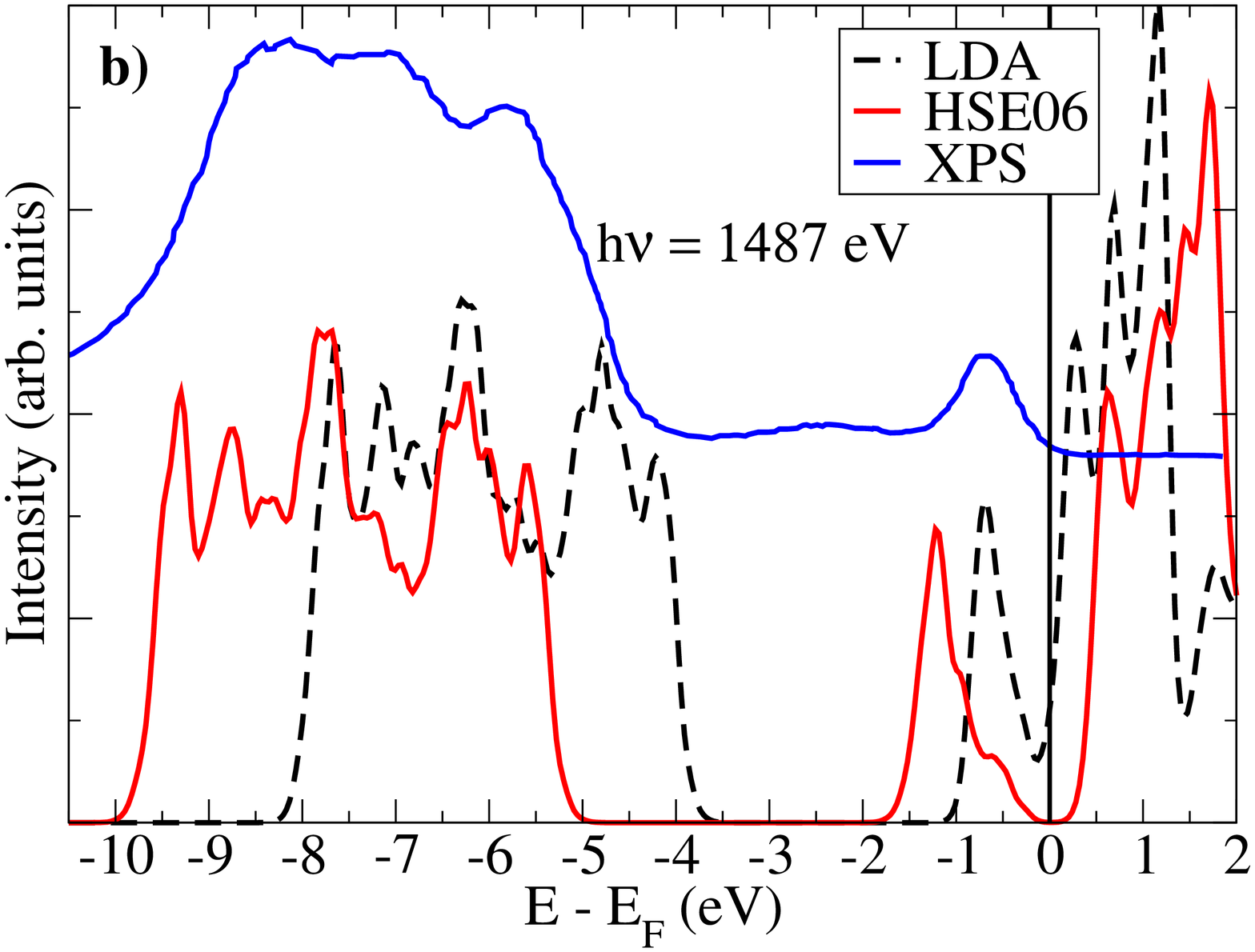}
\caption{(Color online) (a) LDA and HSE06 densities of states (upper and bottom panels, respectively), together with the DOS projected onto Ti $s$, Ti $d$ and O $p$ states, for insulating Ti$_2$O$_3$ at room temperature. (b) Comparison of the total DOS in LDA and HSE06 with the XPS spectrum from Ref. \protect\cite{koethe_phd}.}
\label{figti2o3_1}
\end{center}
\end{figure}

\begin{figure}
\begin{center}
\includegraphics[width=\columnwidth]{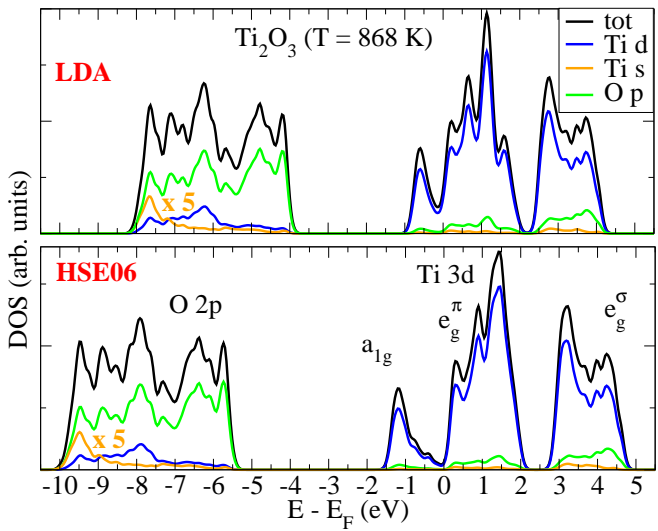}
\caption{(Color online) Comparison between LDA and HSE06 DOS and Ti $s$, Ti $d$ and O $p$ projected DOS, calculated for metallic Ti$_2$O$_3$ at T = 868 K.}
\label{figti2o3_2}
\end{center}
\end{figure}

The effect of the HSE06 corrections over LDA  [see Figs. \ref{figti2o3_1}(a) and \ref{figti2o3_2}] is visible for both phases: i) at the Fermi energy, 
where the top valence band is split off from the bottom of the conduction states, leading to the gap opening in the insulator and to a strong reduction of the spectral weight at the Fermi energy in the metal; ii) in the transfer of spectral weight in the unoccupied Ti $t_{2g}$ band towards the high-energy part of the band; 
iii) in the increasing of the separation between Ti $d$ states and O $p$ states, as a consequence of a rigid shift of the latter. In general, HSE06 results do not change the hybridization character of the bands with respect to LDA.
 
At room temperature HSE06 gives a 0.57 eV gap,  which is much larger than the 0.11 eV estimate from conductivity and thermoelectric coefficient measurements\cite{shin73}. However, contrary to HF calculations\cite{catti97}, spin-polarized HSE06 correctly yields a non-magnetic solution also in the insulating phase. 
This result is important since it shows that within the hybrid functional the existence of a gap is not necessarily linked to the presence of magnetic order, as it often happens in LDA+U. Moreover, HSE06 correctly describes also the metallic phase and the MIT (see Fig.\ref{figti2o3_2}), which is often problematic within LDA+U.  Thus, these results appear to be two advantages of HSE06 with respect to LDA+U in non-magnetic insulators and metals. 

In the XPS spectrum [at h$\nu$ = 1487 eV, reproduced in Fig. \ref{figti2o3_1}(b)] \cite{koethe_phd} for the insulating phase at room temperature
a satellite is clearly visible at 2.4 eV, between the Ti 3$d$ peak at 0.7 eV and the broad O 2$p$ band at 4-10 eV. 
This satellite cannot be obtained with the static hybrid functional employed here, but is absent also in the cluster DMFT calculation\cite{poteryaev03}. 
The HAXPES spectrum (h$\nu$ = 5931 eV) \cite{koethe_phd}, which is dominated by the cation $s$ contribution as in the vanadium oxides \cite{gatti11,papalazarou09} (see Fig. \ref{figv2o3_2}), confirms that the satellite present in XPS spectrum is a genuine bulk feature of insulating Ti$_2$O$_3$ (instead, experimental photoemission results for the metallic phase are not available). Overall, the HSE06 results [see Fig. \ref{figti2o3_1}(b)] compare much better with the experimental spectra than LDA.

\subsection{LaTiO$_3$ and YTiO$_3$}

In a seminal paper, Fujimori {\it et al.} \cite{fujimori92} considered a series of $d^1$ transition metal oxides. 
On the basis of a single-band Hubbard model, they explained the opening of the band gap, going from metallic VO$_2$ and SrVO$_3$ to insulating YTiO$_3$ and LaTiO$_3$, 
as the progressive increase of the ratio between the Hubbard U and the $d$ bandwidth, which is accompanied by the transfer of spectral weight from the quasiparticle peak at the Fermi energy in the metals to the Hubbard bands in the insulators. 
Within this view, both  LaTiO$_3$ and YTiO$_3$ are Mott insulators, with a gap opening between the lower and the upper Hubbard bands.
Thus a band-structure description would not be able to reproduce these incoherent atomic-like excitations in the spectra, and hence the gap.
In fact, LDA KS band structures are metallic for both compounds \cite{solovyev96,sawada98}, while LDA+U \cite{solovyev96,sawada98,okatov05} 
and LDA+DMFT  \cite{pavarini04,craco06,craco07} calculations correctly yield a gap.

At low temperatures both compounds order magnetically. Below 148 K LaTiO$_3$ displays a G-type antiferromagnetic order with a local magnetic moment of 0.57 $\mu_B$ \cite{cwik03}. YTiO$_3$ is ferromagnetic below 29 K, where the magnetic moment is 0.8 $\mu_B$ per Ti atom \cite{knafo09}. 
It has been much debated  whether the magnetic properties of these compounds can be explained in terms of the formation of an orbital liquid or orbital ordering \cite{khaliullin00,haverkort05,mochizuki04}.

Here we use the low-temperature experimental crystal structures from Refs. \cite{cwik03,komarek07}, and a 4$\times$4$\times$4 grid of $\bfk$ points.
While LaTiO$_3$ is non-magnetic in LDA, in HSE06 the local magnetic moment is 0.76 $\mu_B$, overestimating the experimental value of 0.57 $\mu_B$.
Also in ferromagnetic YTiO$_3$, in HSE06  it increases up to 0.84 $\mu_B$  from 0.7 $\mu_B$ in LDA, reaching a similar value than in GGA+U \cite{okatov05}.

\begin{figure}
\begin{center}
\includegraphics[width=0.9\columnwidth]{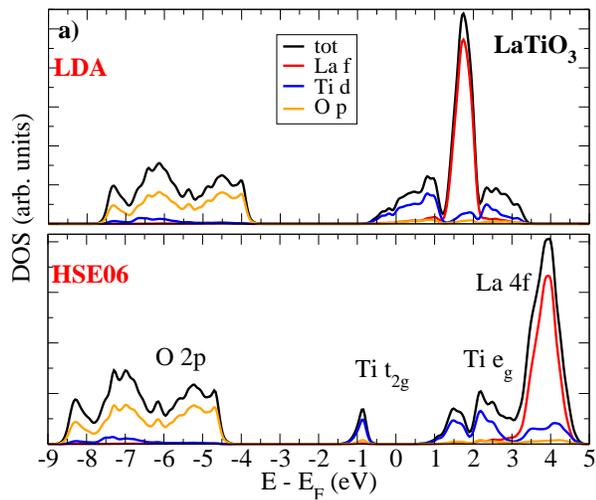}\\[1cm]
\includegraphics[width=0.9\columnwidth]{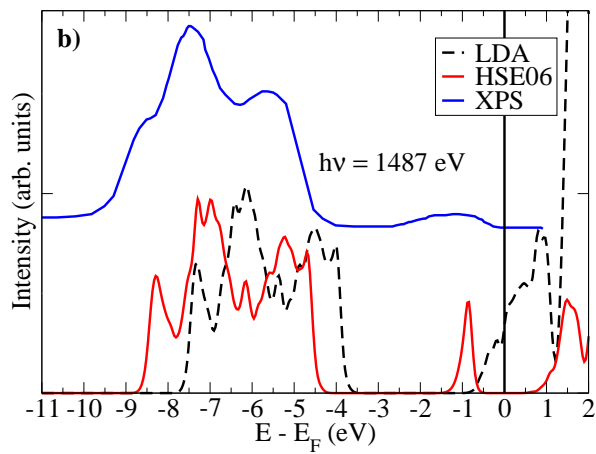}
\caption{(Color online) (a) DOS and projected DOS for LaTiO$_3$ calculated in LDA and HSE06 and (b) comparison with experimental photoemission spectra from Ref. \protect\cite{roth_phd}.}
\label{figlatio3_1}
\end{center}
\end{figure}

\begin{figure}
\begin{center}
\includegraphics[width=0.9\columnwidth]{fig8a}\\[1cm]
\includegraphics[width=0.9\columnwidth]{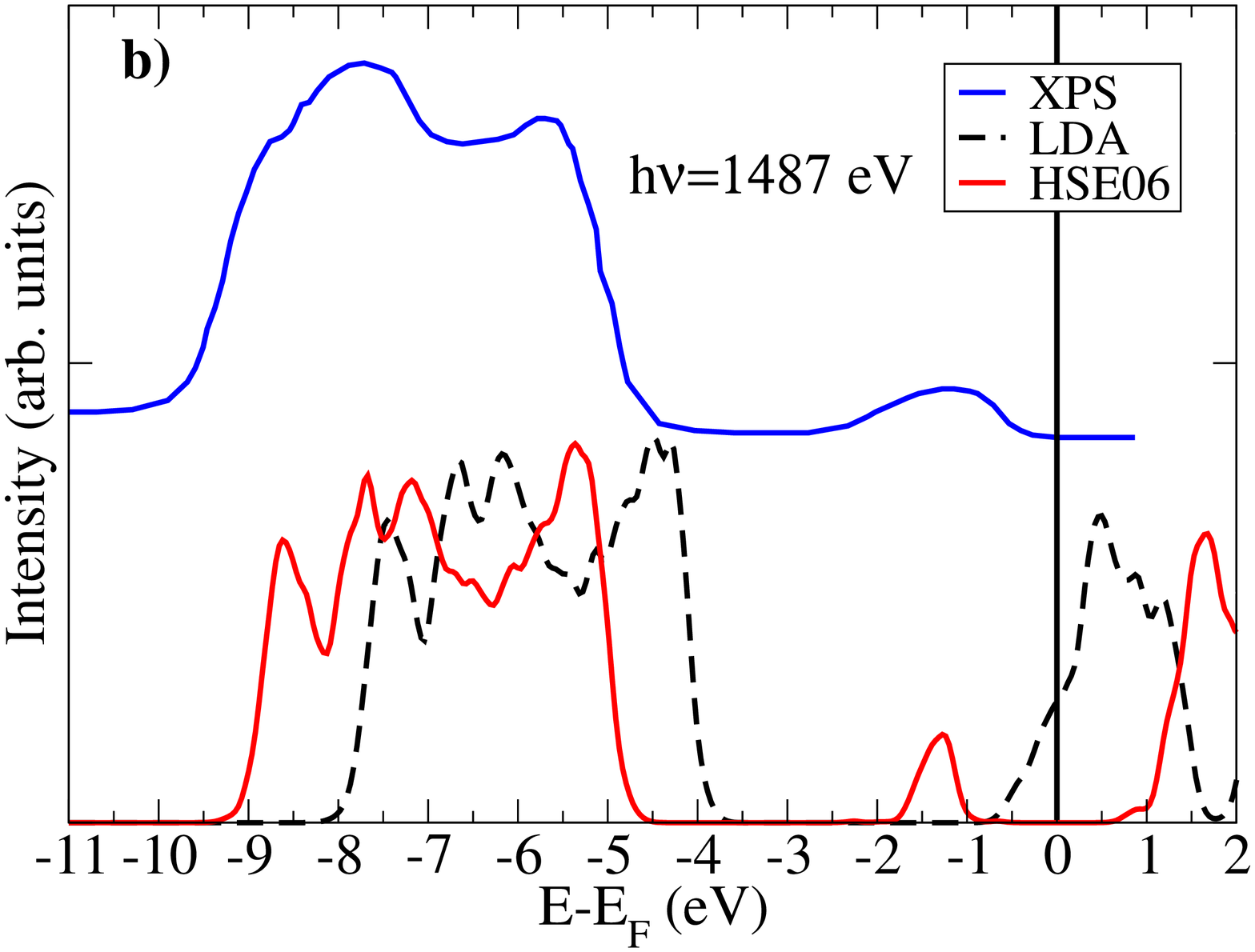}
\caption{(Color online) (a) DOS and projected DOS for YTiO$_3$ calculated in LDA and HSE06 and (b) comparison with experimental photoemission spectra from Ref. \protect\cite{roth_phd}.}
\label{figytio3_1}
\end{center}
\end{figure}

Also in these compounds, the HSE06 gives insulating densities of states [see Figs. \ref{figlatio3_1}(a) and \ref{figytio3_1}(a) for LaTiO$_3$ and YTiO$_3$, respectively]. This is the result of the splitting off of a Ti $t_{2g}$ band from the states crossing the Fermi level in LDA.  In the case of YTiO$_3$ only spin up states contribute to the occupied $t_{2g}$ band. 
Thus, these results for both compounds seem to be in contrast with the traditional interpretation of the topmost occupied state as an incoherent lower Hubbard band \cite{fujimori92}. In fact, the peak in the experimental spectra \cite{roth_phd} [see Figs. \ref{figlatio3_1}(b) and \ref{figytio3_1}(b)] is matched, at least partially, by this (coherent) Ti $t_{2g}$ band. Moreover, within the hybrid functional scheme, the opening of the gap with respect to metallic LDA DOS is due to non-local exchange.
However, contrary to the experiment, in HSE06 the band width of these valence Ti $t_{2g}$ states is larger in YTiO$_3$ than in LaTiO$_3$.
At the same time, also the band gap is larger in LaTiO$_3$ (1.74 eV) than in YTiO$_3$ (1.41 eV),
while the experimental optical gap  is 0.2 eV for LaTiO$_3$ \cite{okimoto95} and  0.7 eV for YTiO$_3$\cite{loa07}.
In both compounds the band gap is overestimated, while the occupied Ti $t_{2g}$ band width is underestimated, also for possible dynamical effects\cite{pavarini04} that are missing here.

While, as in all the other compounds, the hybridization between O $p$ states and Ti $d$ states does not change much between LDA and HSE06 [see Figs. \ref{figlatio3_1}(a) and \ref{figytio3_1}(a)],  a larger effect is seen here for the unoccupied La $f$ states in LaTiO$_3$. 
In LDA they are located in the middle of the conduction band, while in HSE06 they are shifted to the upper end of the band.
The band gap opening between Ti $d$ states and the upshift of the La $f$ states can been obtained in the GGA+U approach\cite{okatov05} 
only with the simultaneous use of two (different) Hubbard U values applied to  Ti $d$ and La $f$ states. 
Within hybrid functionals this result emerges naturally as a consequence of the localization of these states (the non-local exchange
corrects the LDA delocalization error treating all the electrons on equal footing).

Both LaTiO$_3$ and YTiO$_3$ remain insulating also above the corresponding (N\'eel or Curie) temperatures where they loose their magnetic order. 
A spin-unpolarized HSE06 calculation would not be able to obtain a gap in this case. 
In agreement with Mott picture \cite{mott74}, while the  long-range magnetic order is not essential to have an insulator, 
the electronic spins  do matter. In fact, LaTiO$_3$ and YTiO$_3$ in the disordered phases are both paramagnetic.
Similarly, above the N\'eel temperatures transition-metal monoxides are also paramagnetic insulators. 
For these compounds it was recently shown that a calculation, based on the self-interaction-corrected (SIC)  functional 
and taking explicitly into account the disordered local moments of the paramagnetic phase, was able to correctly describe the insulating phases \cite{hughes08}. 
Thus, a similar calculation would be suitable in the (disordered) paramagnetic phase also for the present perovskite compounds (for which a modified SIC implementation has been recently used for the magnetically ordered phases \cite{filippetti11}).

\section{Conclusions}
\label{sec4}

The hybrid functionals  employed in the present work are not explicitly designed  to treat electronic correlations.
Nevertheless, we have shown that the inclusion of non-local Fock exchange is essential to cure the fundamental problem of getting metallic band structures in Kohn-Sham LDA in the insulating phases of several correlated transition-metal oxides, as those that have been discussed here: VO$_2$, V$_2$O$_3$, Ti$_2$O$_3$, LaTiO$_3$, and YTiO$_3$. Analogous results have been obtained for instance also by R\"odl {\it et al.} \cite{roedl09} in the series of transition-metal monoxides, where hybrid functionals were also used as an improved starting point for one-shot GW calculations \cite{fuchs07}.

Thus, a common conclusion emerges from the study of all these correlated transition-metal oxides using HSE06.
With respect to LDA, besides providing a finite band gap, they also correct the position of the O $p$ states. 
This is a clear advantage with respect to other approaches (LDA+U, LDA+DMFT, etc.) stemming from treating all the electrons on the same footing.
Moreover, another advantage with respect to LDA+U is the consistent treatment of insulators and metals and the fact that hybrid functionals 
do not yield  an insulator together with magnetic long-range order, as it is often the case in LDA+U.
However, the HSE06 parametrization overestimates the band gap in all the compounds considered here. 
And, in general, the results depend on the choice of the parameters (as in LDA+U and LDA+DMFT) used to build the functional.
Moreover, hybrid functionals miss completely dynamical correlation effects, which are essential for the description of satellites in photoemission spectra \cite{guzzo11}.
Therefore, it is evident that they cannot be (and they are not meant to be) the final answer for the description of spectral properties of correlated transition-metal oxides. However, also in these compounds,  they demonstrate to be very useful for the discussion of the role of non-local exchange, upon which a clean analysis of the effects of (dynamical) electronic correlation can be then established.

\acknowledgments{We thank Lucia Reining for many fruitful discussions. 
Financial support was provided by  Spanish MEC (FIS2011-65702-C02-01 and PIB2010US-00652), 
ACI-Promociona (ACI2009-1036), Grupos Consolidados UPV/EHU del Gobierno Vasco (IT-319-07), 
and the European  Research Council Advanced Grant DYNamo (ERC-2010-AdG -Proposal No. 267374). 
Computational time was granted by i2basque and BSC ``Red Espanola de Supercomputacion".}

\end{document}